\documentclass[journal,twoside,web]{ieeecolor}
\usepackage{tmi}
\usepackage{cite}
\usepackage{amsmath,amssymb,amsfonts}
\usepackage{algorithmic}
\usepackage{graphicx}
\usepackage{textcomp}
\usepackage{multirow}
\usepackage{bm}
\usepackage[english]{babel}
\newtheorem{theorem}{Theorem}
\usepackage{subfig}
\usepackage{hyperref}

\newcommand{\bqn}{\begin{eqnarray}}
\newcommand{\eqn}{\end{eqnarray}}
\newcommand{\bq}{\begin{eqnarray*}}
\newcommand{\eq}{\end{eqnarray*}}

\makeatletter
\renewcommand*\env@matrix[1][c]{\hskip -\arraycolsep
	\let\@ifnextchar\new@ifnextchar
	\array{*\c@MaxMatrixCols #1}}
\makeatother

\def\BibTeX{{\rm B\kern-.05em{\sc i\kern-.025em b}\kern-.08em
    T\kern-.1667em\lower.7ex\hbox{E}\kern-.125emX}}
\begin{document}
\title{Hodge Laplacian of Brain Networks} 
\author{D. V. Anand and Moo K. Chung
\thanks{This paragraph of the first footnote will contain the date on which
you submitted your paper for review. This study is funded by NIHR01 EB022856, EB02875, NSF MDS-2010778. Correspondence: Moo K. Chung. }
\thanks{D. V. Anand and Moo K. Chung are with the Department of
	Biostatistics and Medical Informatics, University of Wisconsin–Madison,
	Madison, WI 53706 USA (e-mail: mkchung@wisc.edu).}
}
\maketitle

\begin{abstract}
The closed loops or cycles in a brain network embeds higher order signal transmission paths, which provide fundamental insights into the functioning of the brain. In this work, we propose an efficient algorithm for systematic identification and modeling of cycles using persistent homology and the Hodge Laplacian.  Various statistical inference procedures  on cycles are developed.
We validate the our methods on simulations and apply to brain networks obtained through the resting state functional magnetic resonance imaging. {  The computer codes for the Hodge Laplacian 
are given in  \url{https://github.com/laplcebeltrami/hodge}.}

\end{abstract}

\begin{IEEEkeywords}
Hodge Laplacian, Wasserstein distance, brain networks, Cycle basis, Heat kernel smoothing
\end{IEEEkeywords}

\section{Introduction}
\label{sec:introduction}
Understanding the collective dynamics of brain networks has been a long standing question and continues to remain elusive. Many symptoms of the brain diseases such as schizophrenia, epilepsy, autism, and Alzheimer's disease (AD) have shown possible connections with abnormally high levels of synchrony in neural activity \cite{uhlhaas2006neural}. 
The mechanisms underlying the emergence of this synchronous behaviour, is often attributed to the higher order interactions that occur at multiple topological scales \cite{park2013structural,betzel2017multi}. The higher order interactions are evidenced across multiple spatial scales in neuroscience such as collective firing of neurons \cite{uhlhaas2006neural}, 
simultaneous activation of multiple brain regions during cognitive tasks \cite{pessoa2014understanding}. The consideration of higher-order interactions can be highly informative for understanding neuronal synchronisation and co-activation of brain areas at different scales of the network \cite{ghorbanchian2021higher}.

Over the past several decades, significant progress has been made in understanding the structural and functional behavior of the human brain using functional magnetic resonance images (fMRI). In typical fMRI network studies, the brain is usually modelled as a graph whose nodes are specific brain regions and their connectivity is determined by the strength of dependency between the brain regions. Often graph theory based methods have been applied to analyze the brain networks using  quantitative measures such as centrality, modularity and  small-worldness \cite{bullmore2009complex,sporns2018graph,rubinov2011weight,bassett2017small}, which allows to interpret and understand the spatial and functional organization of the brain. Besides, graph measures also provide reliable and quantifiable biomarkers that can discriminate normal and clinical populations \cite{rubinov2010complex}. Hence, the graph measures are used to identify and quantify the differences in the functional networks at both the individual and group level \cite{bullmore2009complex}. The graph comparisons are often performed in the form of either distance-based comparisons or statistics applied to graph theory features \cite{farahani2019application,mheich2020brain,bullmore2009complex,chung2017topological}.

Although graph-based methods can be used to identify graph attributes at disparate scales ranging from local scales at the node level up to global scales at the community level, their power is limited to mostly pairwise dyadic relations \cite{sporns2018graph}. The inherent {\em dyadic} assumption limits the types of neural structure and function that the graphs can model \cite{giusti2016two,battiston.2020}. Therefore, brain network models built on top of graphs cannot encode higher order interactions, i.e., three- and four-way interactions, beyond pairwise connectivity  {\em without} additional analysis \cite{skardal.2020}. To overcome these limitations, we propose to use topological data analysis (TDA), which has gained a lot of traction in recent years due to its simplistic construct in systematically extracting information from hierarchical layers of abstraction. The algebraic topology in TDA has mathematical ingredients that can effectively manipulate structures with higher order relations.  One such a tool is the {\em simplicial complex} which captures many body interactions in complex networks using basic building blocks called simplices \cite{giusti2016two}. The simplicial complex representation easily encode higher order interactions by the inclusion of 2-simplices (faces consisting of 3 nodes) and 3-simplices (volumes consisting of 4 nodes) to graphs. We can further adaptively increase the complexity of connectivity hierarchically from simple node-to-node interaction to more complex higher order connectivity patterns easily. Simplicial complexes have been  used to represent and analyse the brain data \cite{giusti2016two,reimann2017cliques,petri2014homological}. The modular structure of network can easily be recognized by means of connected components, which is the first topological invariant that characterizes the shape of the network. The cycle on the other hand is a second topological invariant which are loops in the network  \cite{chung2019exact,lee2014hole,lee2011computing}.  

Persistent homology (PH), main TDA technique deeply rooted in simplicial complexes, enables network representation at different spatial resolution and provides a coherent framework for obtaining higher order topological features \cite{edelsbrunner2008persistent}. The PH based approaches are becoming increasingly popular to understand the brain imaging data \cite{chung2019exact,sizemore2019importance,chung2017topological}. 
The main approach of PH applied to brain networks is to generate a series of nested networks over every possible parameter through a filtration \cite{petri2013topological}. In particular, the {\em graph filtration} is the most often used filtration specifically designed to uncover the hierarchical structure of the brain networks in a sequential manner \cite{lee2011computing}. 

 Topology-based comparison methods infer the similarity and dissimilarity of networks based on PH feature summaries such as persistent diagrams and persistent landscapes  \cite{songdechakraiwut2020topological,chung2019statistical,chung2019exact}. Typically, a topological discriminating function acts on these PH summaries to discern their topological similarity or dissimilarity \cite{songdechakraiwut2020topological,chung2019statistical,chung2019exact,edelsbrunner2008persistent}. The common topological distances for comparing brain networks are the Gromov-Hausdorff (GH)  and bottleneck (BN) distances \cite{edelsbrunner2008persistent,chung2017topological,chung.2019.NN}.

In the last two decades, PH techniques have made significant inroads in neuroimaging analysis particularly for uncovering global topological features beyond pairwise interactions \cite{sizemore2019importance}. These global features are the topological invariants such as the number of connected components or cycles in a network \cite{lee2012persistent,lee2018abnormal}. Traditional PH based methodologies in neuroimaging have mostly focussed on using these topological invariants as biomarkers for identifying and characterising the topological disparities between the control and diseased populations \cite{lee2014hole,chung.2015.TMI}. While the connected structures of the brain network have been extensively investigated, the studies on the cycles in modeling brain networks is very limited \cite{park2013structural,lee2011computing,lee2014hole,sporns2018graph}. The presence of more cycles in a network signifies dense connections with stronger  redundant connectivity. The cycles in the brain network not only determines the propagation of information but also controls the feedback \cite{lind2005cycles,chung2019statistical}. Since the information transfer through cycles can occur in two different paths, they are sometimes interpreted as redundant connections. Further, cycles are also associated with the information diffusion, dissemination and information bottleneck problems \cite{lee.2014.MICCAI,sizemore.2018,chung.2019.ISBI}. 

While cycles appear naturally in networks, it is not easy to extract or enumerate them.  Cycles are often computed using brute-force depth-first search algorithms \cite{tarjan1972depth}. Recently, a scalable algorithm for computing the number of cycles in the network was proposed \cite{chung2019statistical}. Cycles are usually identified by manipulating  the boundary matrix in PH \cite{edelsbrunner2008persistent,chen2010}. A better approach to determine cycles is by computing the eigenvectors corresponding to zero eigenvalues of the Hodge Laplacian \cite{lee2014hole}. This approach generalizes the graph Laplacian, the 0-th Hodge Laplacian, applied to nodes (0-simplices) to higher order simplexes. Although these algorithms are useful to extract cycles in small networks, it is computationally not feasible to construct and manipulate higher order simplices and extract cycles for large networks. Ideally, we need algorithms that can capture higher order interactions and yet retain the simplicity of graph-based approaches. 

We propose a new spectral method using the Hodge Laplacian that can explicitly identify the connections associated with the cycles. The method is further capable of localizing the connections contributing to the most discriminative cycles in  networks. This is made possible by computing the independent cycle basis and then subsequently building a new topological inference framework that identifies the most discriminating cycles. For the numerical implementation, we propose an efficient new algorithm based on the birth-death decomposition of graphs \cite{songdechakraiwut2020topological}.

\section{Method}
\label{Sec:Methods}
The detailed explanations on TDA tools such as simplicial complexes, birth-death decomposition, Hodge Laplacian over simplicial complexes and the algebraic representation of cycles are presented. \\ 
\subsection{Graphs as a simplicial complex} 
\subsubsection{Simplicial complex}
Consider an undirected complete graph $G = (V, w)$ with vertex set $V$ and edge weight matrix $w=(w_{ij})$ \cite{bullmore2009complex,chung2017topological}. We assume there are $p$ number of nodes.  A binary graph $G_{\epsilon} = (V, w_{\epsilon})$ is a graph consisting of the node set $V$ and the binary edge weights 
$w_{\epsilon} =(w_{\epsilon,ij})$ given by 
\begin{equation}
w_{\epsilon,ij} =   \begin{cases}
	1  \mbox{  if } w_{ij} > \epsilon,\\
	0  \mbox{ otherwise}.
\end{cases}
\label{eq:case}
\end{equation}
Denote $E_\epsilon$ the edge set consisting of all the edges with nonzero weights. Then we may also represent the binary graph $G_\epsilon$ as $G_\epsilon = (V, E_\epsilon)$  if there is no ambiguity. 

\begin{figure}
	\centering
	\includegraphics[scale=0.7]{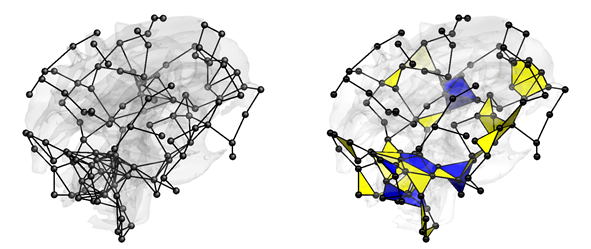}
	\caption{(a) Illustration of brain network representation using a graph (left) and a simplicial complex (right). The graph has only nodes and edges. The simplicial complex  has higher dimensional objects such as triangles (yellow) and tetrahedrons(blue) in addition to nodes and edges.}
	\label{fig:Brain_GF_SC}
\end{figure}

A $p$-simplex $\sigma_p =  [ v_0, v_1,  \cdots , v_p]$ is the convex hull of  $p + 1$ algebraically independent points $v_0, v_1, \cdots, v_p$.
A simplicial complex is a collection of simplices such as nodes ($0-$simplices), edges ($1$-simplices), triangles ($2$-simplices), a tetrahedron ($3$-simplices) and higher dimensional counterparts. A simplicial complex can be viewed as the higher dimensional generalization of a graph \cite{edelsbrunner2008persistent}. Figure \ref{fig:Brain_GF_SC} illustrates the difference between graphs and simplicial complexes in representing a brain network.  

\subsubsection{Chain complex} 
\begin{figure}[t!]
	\centering
	\includegraphics[width=1.0\linewidth]{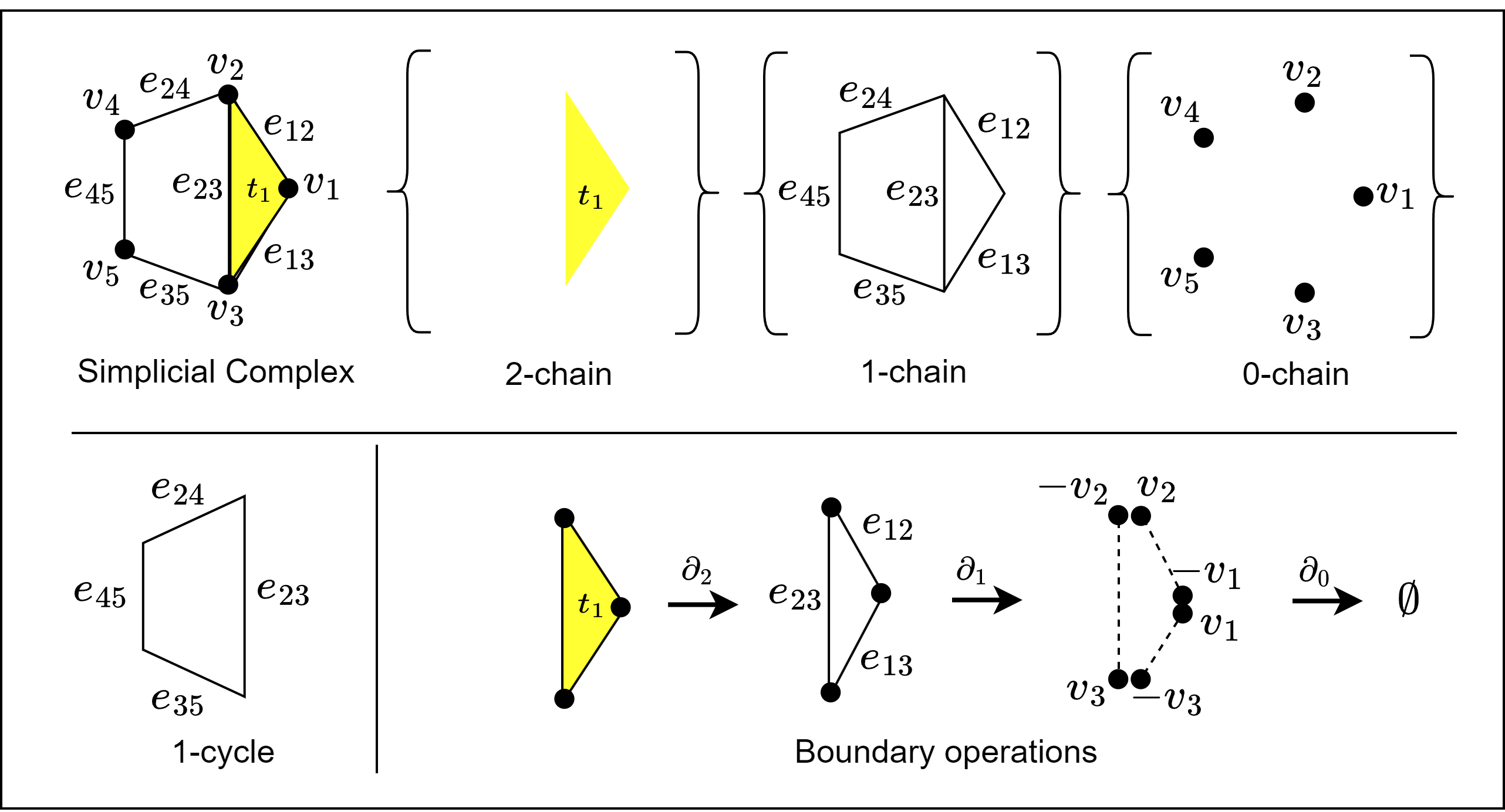}
	\caption{Top left: A simplicial complex with five vertices (0-simplex), six edges (1-simplex) and a triangle (2-simplex). The triangle is represented by  $t_1 = [v_1, v_2, v_3]$ with a filled-in face (colored yellow). Top left: A  chain complex showing 2-chain (set of triangles), 1-chain (set of edges) and 0-chain (set of nodes). Bottom left: The 1-cycle which is present in the simplicial complex. Bottom right: A sequence of boundary operations applied to $t_1$.}
	\label{Fig:SC_boundaryOper}
\end{figure}

{  A $p$-chain is a sum of $p$-simplices denoted as $c = \sum_i \alpha_i\sigma_i,$
where $\sigma_i$ are the $p$-simplices and the $\alpha_i$ are either 0 or 1  \cite{edelsbrunner2010computational}.
The collection of $p$-chains forms a group and the sequence of these groups is called a chain complex. To relate chain groups, we denote a boundary operator $\partial_p  : C_p \rightarrow C_{p-1}$, where $C_p$ denotes the $p$-th chain group.} 
For an oriented $p$-simplex $\sigma_p$ with the ordered vertex set, the boundary operator is defined as 
\begin{equation*}
	\partial_p\sigma_p =  \sum_{i=0}^{p} (-1)^i [v_0, v_1, \cdots , \widehat{v}_i, \cdots , v_p], \label{Eq:orientedsimplex}
\end{equation*}
where $[v_0, v_1, \cdots , \widehat{v}_i, \cdots , v_p]$ is a ($p-1$)-simplex generated from $\sigma_p =  [ v_0, v_1,  \cdots , v_p]$ excluding $\widehat{v}_i$. The boundary operator maps a simplex to its boundaries. Thus, $\partial_2\sigma_2$ maps a triangle to its three edges.
We can algebraically show that  \cite{edelsbrunner2008persistent}
$$\partial_{p-1} \partial_p \sigma_p= 0.$$ 
Figure \ref{Fig:SC_boundaryOper} displays a toy example of a simplicial complex with five vertices (0-simplex), six edges (1-simplex) and a triangle (2-simplex). The triangle is represented by  $t_1 = [v_1, v_2, v_3]$ with a filled-in face (colored yellow). A chain complex showing 2-chain (set of triangles), 1-chain (set of edges) and 0-chain (set of nodes) is shown on the top right. On the bottom left is the 1-cycle present in the simplicial complex and on the bottom right, a sequence of boundary operations is applied to $t_1$. After boundary operation $\partial_2$, we get the $1$-simplices 
$$[v_1, v_2] + [v_2, v_3] - [v_1, v_3] = e_{12} + e_{23} - e_{13},$$ 
which is the boundary of the triangle \cite{lee2014hole,sizemore2019importance}.

\subsubsection{Cycles} A $p$-cycle is a $p$-chain whose boundary is zero. In a graph (1-skeleton), $1$-cycles are loops and $0$-cycles are nodes. To compute $p$-cycles, we use the kernel and image for the boundary operator and establish their relation to the $p$-cycle \cite{edelsbrunner2008persistent,xia2014persistent}. Let $Z_p$ be the collection of all the $p$-cycles given by
$$Z_p = ker \partial_{p} = \{ \sigma_p \in C_p | \partial_{p} \sigma_{p} = 0 \}.$$ 
Let $B_p$ be the boundaries obtained as 
$$B_p = img \partial_{p+1} = \{ \sigma_p \in C_p |  \sigma_p = \partial_{p+1} \sigma_{p+1}, \sigma_{p+1} \in C_{p+1} \}.$$
Since any boundary $\partial_{p+1} \sigma_{p+1}  \in B_p$ satisfies 
$$\partial_p \partial_{p+1} \sigma_{p+1} = 0,$$
it is a $p$-cycle and $B_p \subset Z_p.$ 
Thus, we can partition $Z_p$ into cycles that differ from each other by boundaries through the quotient space
$$H_p= Z_p/B_p,$$
which is called the  $p$-th homology group. The $p$-th Betti number $\beta_p$ counts the number of algebraically independent $p$-cycles, i.e.,
$$\beta_p = rank H_p = rank Z_p - rank B_p.$$
In graph $G_{\epsilon}$, which is 1-skeleton, 
Betti numbers $\beta_0(G_{\epsilon})$ and $\beta_1(G_{\epsilon})$ counts the number of connected components (0-cycles) and number of loops (1-cycles) respectively at threshold $\epsilon$. Betti numbers other than $\beta_0$ and $\beta_1$ are all zero in graphs. 

\subsubsection{Birth-death decomposition}
The graph filtration of weighted graph $G$ is defined as a sequence of nested binary networks \cite{chung2017topological,lee2011computing}: 
$$ G_{\epsilon_0} \supset  G_{\epsilon_1}  \supset   \cdots  \supset  G_{\epsilon_k}$$  where $\epsilon_0 < \epsilon_1 < \cdots < \epsilon_k$ are the sorted edge weights called the {\em filtration values}\cite{lee2011computing,chung2017topological}. The birth and death of $k$-cycles during the process of filtration is quantified using {\em persistence}, which is the duration of filtration values from birth to death. The persistence is usually represented as one-dimensional intervals as persistent barcode (PB)  or two-dimensional  scatter points as a persistent diagram (PD) with the $x$-axis representing birth values and the $y$-axis representing death values \cite{edelsbrunner2008persistent}.

During the graph filtration, once a component is born, it does not die. Thus, all the death values of connected components are $\infty$ and can be ignored. Then the total number $\mathcal{P}$  of birth values of connected components (0-cyles) is
\begin{equation}
\mathcal{P} = \beta_0(G_{\infty}) - 1 = p - 1. 
\label{Eq:beta0cnt} 
\end{equation}
The 0D barcode corresponding to 0-cycles consists of a set of increasing birth values 
$$B(G) = b_1 <  b_2  < \cdots < b_{\mathcal{P}}.$$
During the filtration, cycle is considered as  born at $-\infty$. All the birth values of 1-cycles can be ignored.  
The 1D barcode corresponding to 1-cycles consists of a set of increasing death values 
$$D(G) = d_1 <  d_2  < \cdots < d_{\mathcal{Q}}.$$
For a graph with $q=p(p-1)/2$ number of edges,  the total number of edges $q$ is equivalent to
$q = \mathcal{P} + \mathcal{Q}.$
Thus, we have $\mathcal{Q} =  (p - 1)(p - 2)/2$ number of death values of 1-cycles. During the filtration, the birth of a component and the death of a cycle cannot occur at the same instant and this can be more formally stated as \cite{song.2021.MICCAI}:

\begin{theorem}[Birth-death decomposition]
	\label{Th:bddecomp}
	The set of 0D birth values $B(G)$ and 1D death values $D(G)$ partition the edge weight set $W$ such that 
	$$W = B(G)\cup D(G), \quad B(G)\cap D(G) = \emptyset.$$ The cardinalities of $B(G)$ and $D(G)$ are $p -1$ and $(p-1)(p-2)/2$ respectively. 
\end{theorem}

In the graph filtration, the birth values are easily computed using the maximum spanning tree (MST). Given a weighted graph $G$, computing the set of 0D birth values $B(G)$ is equivalent to the finding MST of $G$ through 
Kruskal’s or Prim’s algorithms \cite{lee2011computing,songdechakraiwut2020topological}.  Once $B(G)$ is computed,$D(G)$ is simply given as the rest of the remaining edge weights that are not part of MST. Thus, the barcodes for 0- and 1-cycles can be computed efficiently in $\mathcal{O}(q \log p)$.

\subsubsection{Wasserstein distance on 1-cycles}
\label{Sec:WSDGraph}
The topological similarity or dissimilarity between the networks can be inferred from the differences between barcodes \cite{mileyko2011probability}. The Wasserstein distance is a metric that is often used to quantify the underlying differences in the barcodes \cite{mi.2018,mi.2020,song.2021.MICCAI}.  Let $\Omega = (V^\Omega, w^\Omega)$ and $\Psi = (V^\Psi, w^\Psi)$ be two given networks with $p$ nodes. Their persistent diagrams denoted as $P_{\Omega}$ and $P_{\Psi}$ are expressed in terms of scatter points as
\bq x_1 &=& (b_1^{\Omega}, d_1^{\Omega}), \cdots, x_q = (b_\mathcal{Q}^{\Omega},d_\mathcal{Q}^{\Omega})\\
y_1 &=& (b_1^{\Psi}, d_1^{\Psi}), \cdots, y_q = (b_\mathcal{Q}^{\Psi}, d_\mathcal{Q}^{\Psi})
\eq
respectively. We can show that the $2$-{\em Wasserstein distance} on persistent diagrams  is given by
$$\mathcal{D}(P_{\Omega}, P_{\Psi}) = \inf_{\tau: P_{\Omega} \to P_{\Psi}} \Big( \sum_{x \in P_{\Omega}} \| x - \tau(x) \|^2 \Big)^{1/2}$$
over every possible bijection $\tau$ between $P_{\Omega}$ and $P_{\Psi}$ \cite{song.2021.MICCAI}. 
Since persistent diagrams are 1D scatter points for graph filtrations, the bijection $\tau$  is simply given by matching sorted scatter points \cite{song.2021.MICCAI}:

\begin{theorem} 
	The 2-Wasserstein distance between the 1D persistent diagrams (1-cycles) for graph filtrations is given by 
	$$\mathcal{D}_{1}(P_{\Omega}, P_{\Psi}) =  \Big[ \sum_{i=1}^{q} (d_{(i)}^{\Omega} - d_{(i)}^{\Psi})^2  \Big]^{1/2},$$
	where $d_{(i)}^{\Omega}$ and $d_{(i)}^{\Psi}$ are the $i$-th smallest death values associated with 1-cycles (loops). 
	\label{theorem:optimal}
\end{theorem}

\subsection{Hodge Laplacian over simplicial complexes}

The Hodge Laplacian generalizes the usual graph Laplacian for nodes (0-simplices) to $p$-simplices. The Laplacian matrix $\mathcal{L}_0$  for a graph is given by $ \mathcal{L}_0 = D - A$ with degree matrix $D$ and adjacency matrix $A$. In general, a higher-dimensional Laplacian can be defined for each dimension $p$ using two matrices that perform the role of upper and lower adjacency matrices: 
$$ \mathcal{L}_p = \mathcal{L}_p^U + \mathcal{L}_p^L $$
where $\mathcal{L}_p^U$ and $\mathcal{L}_p^L$ are the upper and lower adjacency Laplacians \cite{battiston.2020}. 

\subsubsection{Hodge Laplacian} The higher dimensional Laplacian $\mathcal{L}_p$ is usually referred to as the Hodge Laplacian or the $p$-Laplacian that connects the $p$-simplices with their adjacent $(p+1)$-(upper adjacency) and $(p-1)$-simplices (lower adjacency). Consider boundary matrix $\mathcal{B}_p$ representing boundary operator $\partial_p$ \cite{meng2021persistent}
\begin{equation}
	(\mathcal{B}_p)_{ij} =
	\begin{cases}
		1, 	& \text{if } \sigma^i_{p-1}  \subset \sigma^j_{p}  ~~\text{and}~~ \sigma^i_{p-1} \sim \sigma^j_{k}\\
		-1, & \text{if } \sigma^i_{p-1}  \subset \sigma^j_{p}  ~~\text{and}~~ \sigma^i_{p-1} \nsim \sigma^j_{k}\\
		0,  & \text{if } \sigma^i_{p-1}  \not \subset \sigma^j_{p}
	\end{cases},
	\label{Eq:boundarymatrix}
\end{equation}
where $\sigma^i_{p-1}$ is the $i$-th ($p$-1)-simplex and $\sigma^j_p$ is the $j$-th $p$-simplex. Notations $\sim$ and $\nsim$ denote similar (positive) and dissimilar (negative) orientations respectively. Then the $p$-th Hodge Laplacian matrix $\mathcal{L}_p$ is defined as
\begin{equation}
	\mathcal{L}_p =	\mathcal{B}^T_p\mathcal{B}_p + \mathcal{B}_{p+1}\mathcal{B}_{p+1}^T. 
	\label{Eq:HodgeLaplacian}	
\end{equation}
The Hodge Laplacian $\mathcal{L}_p$ can be viewed as the sum of the Laplacians from the lower dimensional simplices \cite{horak2013spectra,barbarossa2020topological,schaub2020random,mukherjee2016random}
$$\mathcal{L}_p^L = \mathcal{B}^T_p\mathcal{B}_p $$ and  upper dimensional simplices 
$$\mathcal{L}_p^U = \mathcal{B}_{p+1}\mathcal{B}_{p+1}^T.$$ Since $\mathcal{B}_{0}=0$, the Hodge Laplacian for a 1-skeleton is $\mathcal{L}_0$ = $\mathcal{B}_1\mathcal{B}^T_1$, which is popularly referred as the graph Laplacian. The boundary matrix $\mathcal{B}_1$ relates how nodes are connected to form edges is commonly referred as incidence matrix in the graph theory.  Since there is only $0$-simplices and $1$-simplices in 
a 1-skeleton, the boundary matrix $\mathcal{B}_{2}=0$. Thus, the second term in the Hodge Laplacian $\mathcal{L}_1$ vanishes and we have  $$\mathcal{L}_1 = \mathcal{L}_1^L = \mathcal{B}^T_1\mathcal{B}_1$$
for graphs.

\subsubsection{Algebraic representation of  $1$-cycles}
The spectral decomposition of Hodge Laplacian  is performed to identify $p$-cycles of the underlying network \cite{horak2013spectra,lee2014hole,lim2020hodge,meng2021persistent}. The $p$-th homology group $H_p$ is a kernel of Hodge Laplacian $\mathcal{L}_p$ given by
$$H_p = ker \mathcal{L}_p.$$ 
The eigenvectors with zero eigenvalue of $\mathcal{L}_p$ span the kernel space of $\mathcal{L}_p$. 
We first solve
$$ \mathcal{L}_p {\bf U}_p = \Lambda_{p} {\bf U}_p,$$
where $\Lambda_{p}$ is a diagonal matrix of eigenvalues and ${\bf U}_p$ is a matrix of eigenvectors.
The multiplicity of the zero eigenvalue of Hodge Laplacian $\mathcal{L}_p$ is the Betti number $\beta_p$, the rank of the kernel space of $\mathcal{L}_p$. This is related to the algebraic connectivity and generalizes from the well known fact that the number of zero eigenvalues of the graph Laplacian is the number of connected components. Similarly, the number of zero eigenvalues of the $\mathcal{L}_0$, $\mathcal{L}_1$  and $\mathcal{L}_2$ matrix corresponds to the number of $0$-cycles (connected components), $1$-cycles (closed loops) and $2$-cycles (voids or cavities) respectively. Since the eigenvectors corresponding to the zero eigenvalues are related to the homology generators \cite{friedman1998computing}, 
we represent a 1-cycle using the coefficients of the eigenvectors.  Let $A =(a_{l(i,j),m})$ be the matrix consisting of columns of ${\bf U}_1$ that corresponds to the zero eigenvalue. The size of $A$ is $q \times \beta_1$ with Betti number $\beta_1$.  Here $a_{l(i,j),m}$ is the entries of $m$-th eigenvector of the Hodge Laplacian corresponding to edge $e_{ij}$. The $m$-th $1$-cycle  $\mathcal{C}^m$ is then represented as
\begin{equation}
	\mathcal{C}^m = \sum_{e_{ij} \in E} a_{l(i,j),m} e_{ij}. 
	\label{Eq:kcycle_rep}
\end{equation}
$\mathcal{C}^m$  can be represented as a vector by putting  coefficient $a_{l(i,j),m}$ into the corresponding position in the lexicographically ordered edge set $[e_{12}, e_{13}, \cdots, e_{23}, e_{24}, \cdots, e_{q-1,q}]^T$.

\begin{figure}[t!]
	\centering
	\includegraphics[width=1.0\linewidth]{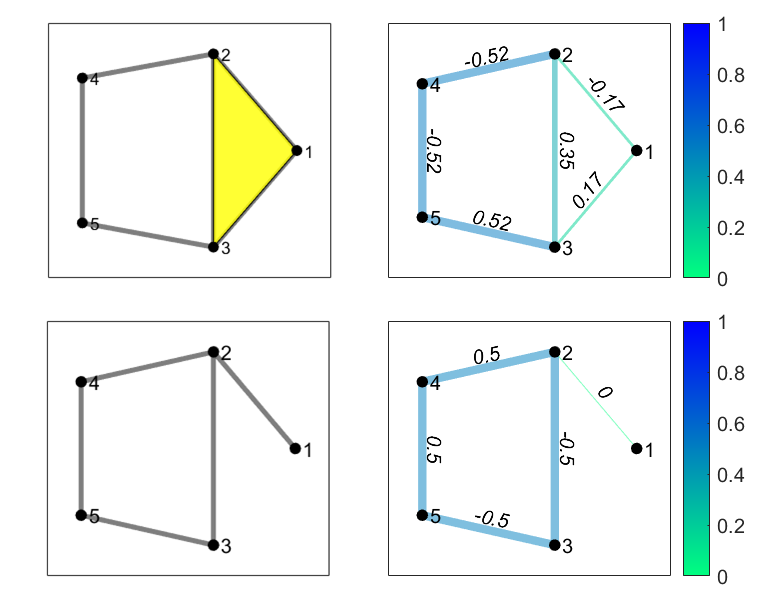}
	\caption{ Top-left: A 2-skeleton representation network made of five vertices connected by six edges. Top-Right: The  $1$-cycle is formed by the vertices $v_2, v_4, v_5, v_3$. The numbers on edges are eigenvector of the Hodge Laplacian $\mathcal{L}_1$ corresponding to the zero eigenvalue. The edge colors indicate the absolute value of coefficients of the cycle representation $C^1$. Bottom-left: edge $e_{13}$ is removed resulting in 1-skeleton. Bottom-right: The 1-cycle identified along with the edges that constitute the cycle.}
	\label{fig:HL_example1}
\end{figure}

For Figure \ref{fig:HL_example1}-bottom example, the eigendecompostion on the Hodge Laplacian $\mathcal{L}_1$ results in the eigenvalues  $\left[0.00, 0.83, 2.00,2.69,4.48\right]$ and the eigenvector corresponding to the zero eigenvalue is obtained as $\left[0.00, 0.50, -0.50, 0.50, -0.50\right]^T$. We ignored $e_{13}$ since there is no connection. The $1$-cycle is then represented as
$$ \mathcal{C}^1 =  0.5e_{23} - 0.5e_{24} + 0.5e_{35} -  0.5e_{45}.$$  
A similar procedure will be used to identify and extract the 1-cycles by breaking down the graph into a subgraph containing only one 1-cycle.

\subsubsection{Computation of 1-cycle basis}
The representation (\ref{Eq:kcycle_rep}) uses  all the edges representing a 1-cycle. Even the edges that are not a part of a cycle are used in the representation. This has been the main limitation of using the Hodge Laplacian in identifying 1-cycles in the past \cite{lee2014hole}. In the proposed method, we split the graph into a series of subgraphs such that each subgraph has only one 1-cycle. 

The graph filtration partitions the edges in a given network uniquely into the birth and death sets. While the edges in the birth set are responsible for creating components, the edges in the death set accounts for destroying cycles. The edges in the birth set  forms the maximums spanning tree (MST) with no cycles. When  adding an edge from the death set to MST, a 1-cycle is formed. The process is repeated sequentially till we use up all the edges in the death set. We claim the resulting 1-cycles form a basis.

\begin{theorem} Let $M(G)$ be the MST of graph $G$. When the $k$-th edge  $d_k$ from the death set $D(G)$ is added to the MST, 1-cycle $\mathcal{C}^k$  is born. The collection of cycles $\mathcal{C}^1, \cdots, \mathcal{C}^Q$ spans $ker \mathcal{L}_p$.
\label{theorem:3}
\end{theorem}
\textit{Proof.} Let $E_k$ be the edge set of the cycle $\mathcal{C}^k$. Since $E_k$ and $E_l$ differ at least by edges $d_k$ and $d_l$, they are algebraically independent. Hence, all the cycles $\mathcal{C}^1, \cdots, \mathcal{C}^{\mathcal{Q}}$ are independent from each other. Since there should be $Q$ number of independent cycles in the 1-st Homology group $H_1 = ker \mathcal{L}_1$, they form a basis. \hfill $\square$

We can sequentially extract 1-cycles using the Hodge Laplacian of the subgraph $G_k = (V, T \cup \{ d_k \})$, which contains only one 1-cycle $\mathcal{C}^k$. We get exactly one eigenvector corresponding to the zero eigenvalue. The entries of eigenvector will be all zero on the edges that are not part of the cycle. Thus, we can represent 1-cycle $\mathcal{C}^k$ only using edges that contribute to the cycle as
\begin{equation}
	\mathcal{C}^k = \sum_{e_{ij} \in E_k} a_{l(i,j),k} e_{ij}. 
	\label{Eq:kcycle_rep2}
\end{equation}
The representation (\ref{Eq:kcycle_rep2}) contains only the edges that form the cycle. All other terms are zero.
Thus, $\mathcal{C}^k$  can be represented as vectors by putting 
$a_{l(i,j),k}$ into the corresponding position in the vectorized edge set $[e_{12}, e_{13}, \cdots, e_{23}, e_{24}, \cdots,$ 
$e_{q-1,q}]^T$. Subsequently, all the 1-cycle basis $\mathcal{C}^1, \cdots, \mathcal{C}^\mathcal{Q}$
can be systematically extracted and efficiently stored as a sparse matrix. Since $\mathcal{C}^1, \cdots, \mathcal{C}^Q$ forms a basis, any cycle in the graph can be represented as a linear combination
$ \sum_{j=1}^Q \alpha_j \mathcal{C}^j.$

\begin{figure}[t]
	\centering
	\includegraphics[width=1.0\linewidth]{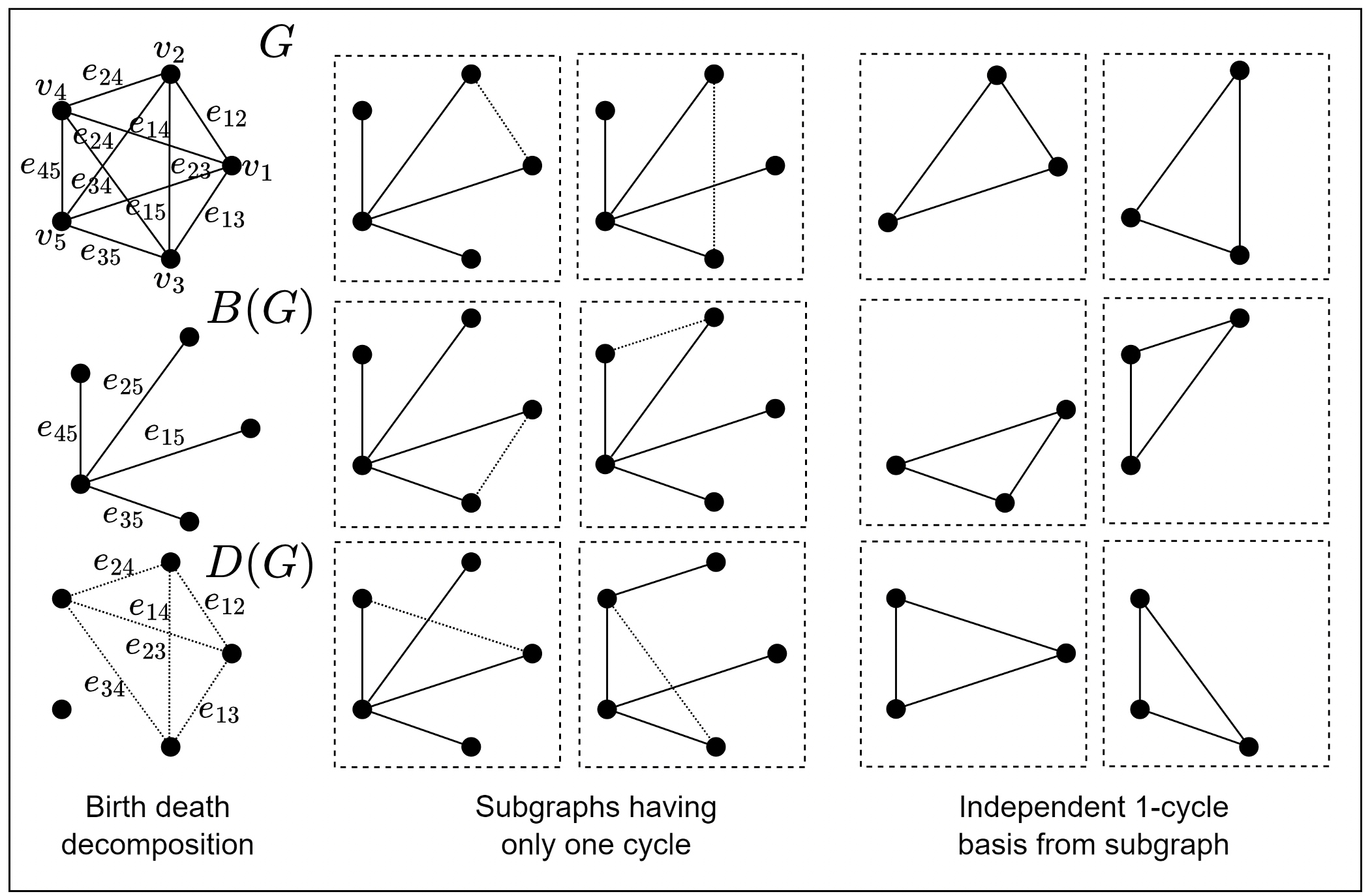}
	\caption{Left: Graph $G$ is decomposed into birth set $B(G)$ with edges $[e_{15}, e_{25}, e_{35}, e_{45} ]$ and death set $D(G)$ with edges $[e_{12}, e_{13}, e_{14}, e_{23}, e_{24}, e_{34}]$. Middle: The subgraphs is constructed by adding an edge from the death set to the birth set. Right: The independent 1-cycles obtained from the eigenvectors of the Hodge Laplacian corresponding to the zero eigenvalue.}
	\label{fig:HodgeLaplacian_kCycle}
\end{figure}

The extraction of 1-cycle basis of a graph can be summarized to three steps (Figure \ref{fig:HodgeLaplacian_kCycle}). {\bf 1)}   The birth-death decomposition is used in extracting birth and death values. The edges $\left[ e_{15}, e_{25}, e_{35}, e_{45} \right]$ form the birth set $B(G)$ and the remaining edges $\left[ e_{12}, e_{13}, e_{14}, e_{23}, e_{24}, e_{34} \right]$ become the death set $D(G)$. The edges in the birth set correspond to the maximum spanning tree (MST). {\bf 2)}  The subgraphs having only one cycle each are created by adding an edge from the death set to the MST. {\bf 3)} The Hodge Laplacian is used in  identifying only edges that belong to each cycle. 

\subsection{Statistical analysis on 1-cycles}

Let $\Omega = \{ \Omega_1, \cdots, \Omega_m \}$ and $\Psi = \{ \Psi_1, \cdots, \Psi_n \}$  be the collections of $m$ and $n$ complete graphs each consisting of $p$ number of nodes. There are exactly $\mathcal{Q} = (p-1)(p-2)/2$ number of cycles in each network. We are interested in developing new statistical inference procedures testing the topological difference between $\Omega$ and $\Psi$.

\subsubsection{Inference on death values} We use the Wasserstein distance between graphs in measuring the 1D topological difference. Consider the average Wasserstein distance within groups $\mathfrak{L}_W$ and between groups $\mathfrak{L}_B$ given by \cite{song.2021.MICCAI}:
\bq \mathfrak{L}_W &=& \frac{1} { {m \choose 2} + {n \choose 2}}    \Big[ \sum_{i<j} \mathcal{D}_1(\Omega_i, \Omega_j) + \sum_{i<j} \mathcal{D}_1(\Psi_i, \Psi_j) \Big]\\
 \mathfrak{L}_B &=& \frac{1}{mn} \sum_{i=1}^m \sum_{i=1}^n \mathcal{D}_1(\Omega_i, \Psi_j).\eq
We are only using the Wasserstein distance between cycles, which are computed using the sorted death values. 
We then use the ratio  $\mathfrak{L}_{B/W} = \mathfrak{L}_{B}/\mathfrak{L}_{W} $ as the test statistic. 
If the two groups are topologically close, $\mathfrak{L}_{B}$ becomes small while $\mathfrak{L}_{W}$ becomes large. Thus the ratio $\mathfrak{L}_{B/W}$ can be used to as test statistic. Since the probability distribution of $\mathfrak{L}_{B/W}$ is unknown, we used the permutation test \cite{chung2019rapid,thompson2001genetic,zalesky2010whole,nichols2002nonparametric,winkler2016faster}. For large sample sizes $m$ and $n$, the permutation test is computationally costly. We adapted for the scalable {\em transposition  test}, which sequentially update the test statistic over transpositions \cite{chung2019rapid,song.2021.MICCAI}. Unlike the permutation test that shuffles all graphs in each permutation, the transposition test only shuffles one graph per group in a permutation.
 Computing the statistic $\mathfrak{L}_{B/W}$ over each permutation requires the recomputation of the Wasserstein distance from scrach. Instead, we perform the transposition of swapping only one graph per group and setting up an iteration of how the test statistic changes over the transposition. In real brain network data,  we used the test statistics with 500000 random transpositions while interjecting a random permutation for every 500 transpositions. The intermix of transpositions and  permutations has the effect of speeding up the convergence \cite{chung2019rapid}.

\subsubsection{Common 1-cycle basis across subjects}
If 1-cycle basis change from one subject to next, it is difficult to use the basis  itself as a feature in statistical analysis. 
Thus, we propose to use a common 1-cycle basis across subjects by using the network template, which is  obtained by averaging correlation matrices of all subjects. Then we encode subject-level variability in the expansion coefficients of the common  1-cycle basis
$\bm{\phi} = \left[\mathcal{C}^1, \mathcal{C}^2, \cdots, \mathcal{C}^{\mathcal{Q}} \right]$
obtained from (\ref{Eq:kcycle_rep2}).
Subsequently, the vectorized upper triangle entries of individual correlation matrix $w$  as  
$$w = \alpha_1 \mathcal{C}^1 + \alpha_2 \mathcal{C}^2 +  \cdots + \alpha_{\mathcal{Q}} \mathcal{C}^{\mathcal{Q}}.$$
The coefficients $ \bm{\alpha} = [\alpha_1, \cdots, \alpha_\mathcal{Q}]^T$ are estimated in the least squares fashion as
\begin{equation}
	\bm{\alpha} = (\bm{\phi}^T\bm{\phi})^{-1} \bm{\phi}^T w.
	\label{Eq:kcyclecoeff}
\end{equation}
The estimated coefficients $\bm{\alpha}$ for each subject are then used in discriminating two groups of networks $\Omega$ and $\Psi$. Let $\bar \alpha_j^\Omega$ and $\bar \alpha_j^\Psi$ be the means of the $j$-th 1-cycle basis in group $\Omega$ and $\Psi$ respectively. Then we used the maximum difference 
\begin{equation}
\mathfrak{L}(\Omega, \Psi) = \max_{1 \leq j \leq  \mathcal{Q}} | \bar \alpha_j^\Omega  - \bar \alpha_j^\Psi |
\label{Eq:coefflossdef}
\end{equation}
as the test statistic in discriminating between the groups. The statistical significance is determined using the permutation test. 
Unlike previous analysis  that cannot localize specific cycles, the test statistic gives a way to localize most discriminating cycles by identifying the $j$-th cycle that gives the maximum.

\section{Validation}
Since cycles can be modeled to embed complex interactions, it can potentially uncover hidden topological patterns which are hitherto impossible in conventional graphical models. In literature, there is no baseline statistical method for explicitly modeling cycles in networks. Also, there is no ground truth in real brain data. Even if we apply the baseline methods to real data, it is unclear which method provides the best answer. Thus, we validated the proposed methods in a simulation study with the ground truth. 
The Matlab codes for the simulation and related Hodge Laplace codes are given in    \url{https://github.com/laplcebeltrami/hodge}.

\begin{figure}[t!]
	\centering
	\includegraphics[width=1\linewidth]{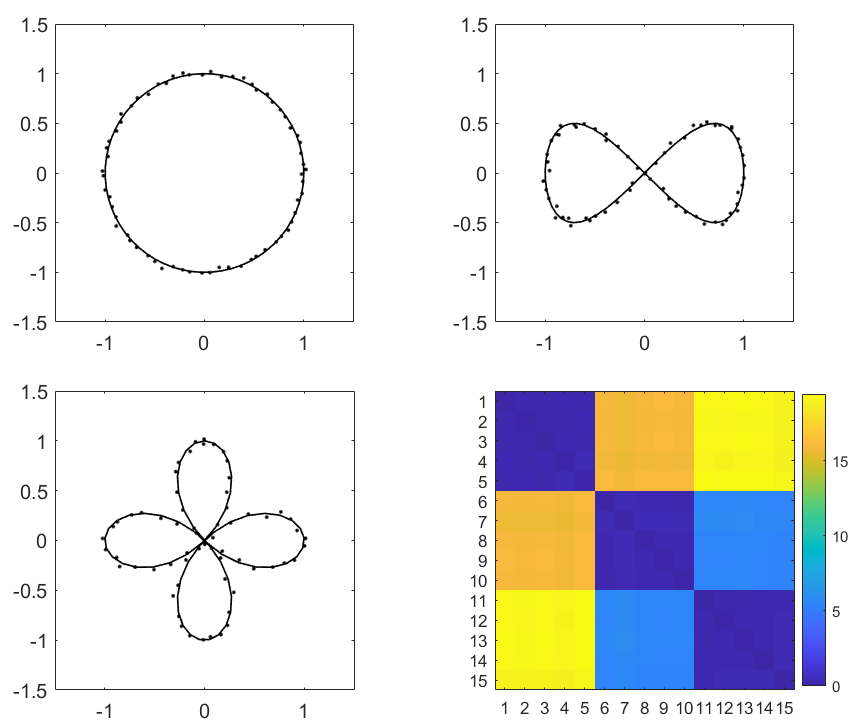}
	\caption{The three types of cycles with different topology: circle (1-loop), lemniscate (2-loops), quadrifolium (4-loops) used in the simulation study. Gaussian noise $N(0, 0.02^2)$ is added to the coordinates. Bottom right: the pairwise Wasserstein distance matrix computed using the death values of the 1-cycles on 5 networks in each group.}
	\label{fig:cycle_wsd123}
\end{figure}

\subsection{Simulation study I} 

We generated three types of networks with different number of loops (1-cycles). Some well known curved shapes such as a circle, leminiscate, quadrifolium \cite{yates1974curves} were chosen as the ground truth and  added Gaussian noise  $N(0, 0.02^2)$ to the coordinates (Figure \ref{fig:cycle_wsd123}). 
The circle has a single loop, the leminiscate has two loops and the quadriform has four loops. The number of nodes to construct the network were chosen as $p=64$  for all the types. This ensures we have the same number of cycles ($Q=1953$ independent 1-cycles) in each type of simulated network. For each type of networks, we generated 6, 8, 10 and 12 number of networks.

	\begin{table}
		\caption{The performance results of the Wasserstein distance on 1-cycles  are summarized as average p-values for testing different loop types. We generated different number of networks. The smaller p-values in the first three rows indicate that our method can discriminate network differences. The larger p-values in the last three rows
		indicate that we are not producing false positives.}
		\label{Table:SimulationStudy1}	
		\renewcommand{\arraystretch}{1.2}
		\begin{tabular}{c|cccc}
			\centering
			loop types & 6 networks & 8 networks & 10 networks & 12 networks\\
			\hline
			1 vs. 2 & $1.4 \times 10^{-3}$ & $6.8 \times 10^{-5}$ & $6.1 \times 10^{-6}$ & $4.4 \times 10^{-7}$\\
			1 vs. 4 & $1.1 \times 10^{-3}$ & $5.4 \times 10^{-5}$ & $4.9 \times 10^{-6}$ & $5.2 \times 10^{-7}$\\
			2 vs. 4 & $1.2 \times 10^{-3}$ & $6.6 \times 10^{-5}$ & $3.2 \times 10^{-6}$ & $2.0 \times 10^{-7}$\\
			\hline \hline
			1 vs. 1 & 0.3954 & 0.5336 & 0.9790 & 0.7834 \\
			2 vs. 2 & 0.6516 & 0.8404 & 0.3458 & 0.5376 \\
			4 vs. 4 & 0.5943 & 0.8294 & 0.7561 & 0.5403 \\
			\hline
		\end{tabular}	
	\end{table}

\subsubsection{Death values} 

The topological distances between the simulated networks were measured by computing the 2-Wasserstein distance between the persistent diagrams of 1-cycles. Figure \ref{fig:cycle_wsd123} shows  the pairwise Wasserstein distance map  showing a clear clustering pattern. Networks with similar topology have smaller distances while networks with different topology have relatively large distances. Using the proposed ratio statistic, we computed p-values comparing different network types. Table \ref{Table:SimulationStudy1} shows the average p-values obtained after 50 independent simulations. Each simulation was perforemd with 100000 permutations. Networks of the same topology have large p-values indicating they are shown to be statistically not different.

When testing networks of different topology (first three rows), we  have small p-values indicating they are shown to be statistically different. The results indicate the proposed method perform well in discriminating networks of different topology. As the number of networks increase in each group, the p-values get smaller showing increased statistical power over increased sample size. When testing networks of identical topology (last three rows), we have large p-values indicating they are not showing statistical differences and  does not produce much false positives. Thus, the method perform well as expected.

\subsubsection{Common 1-cycles basis across subjects} We used the maximum gap between coefficients of 1-cycle basis as the test statistic on the same simulation study.  The test was repeated for 10 times and the average p-values are reported. Each simulation was performed with 100000 permutations. Table \ref{Table:SimulationStudy2a} shows the average p-values obtained for the simulation. The p-values are low for networks with differences while the p-values are large when the network has no difference. The method performed better than Wasserstein distance based method reported in Table \ref{Table:SimulationStudy1}.

\begin{center}
	\begin{table}[t!]
		\caption{The performance results of the common 1-cycle basis method are summarized as the average p-values.  The smaller p-values indicate that our method can discriminate network differences (first three rows) while the larger p-values
		indicate that our method does not produce false positives (last three rows).}
		\label{Table:SimulationStudy2a}	
		\renewcommand{\arraystretch}{1.2}
		\begin{tabular}{c|cccc}
			\centering
			loop-type & 6 networks & 8 networks & 10 networks & 12 networks\\
			\hline
			1 vs. 2 & $2.1 \times 10^{-3}$ & $2.0 \times 10^{-4}$ & $1.0 \times 10^{-5}$ & $0.0000$\\
			1 vs. 4 & $1.9 \times 10^{-3}$ & $1.2 \times 10^{-4}$ & $2.0 \times 10^{-5}$ & $0.0000$\\
			2 vs. 4 & $1.8 \times 10^{-3}$ & $1.4 \times 10^{-4}$ & $1.0 \times 10^{-5}$ & $0.0000$\\
			 \hline \hline
			1 vs. 1 & 0.4263 & 0.6606 & 0.8736 & 0.6735 \\
			2 vs. 2 & 0.3962 & 0.8919 & 0.9620 & 0.5590 \\
			4 vs. 4 & 0.7988 & 0.7365 & 0.4598 & 0.9815 \\
			\hline
		\end{tabular}	
	\end{table}
\end{center}

\subsection{Simulation study II} 

\begin{figure}[t!]
	\centering
	\includegraphics[width=1\linewidth]{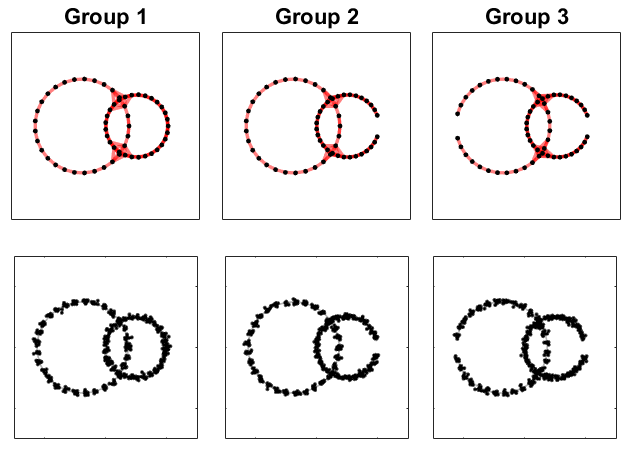}
	\caption{{ Top: Three topologically different network shapes with different number of loops in each group. Group 1 has three loops, Group 2 has two loops and Group 3 has one loop.
	 Bottom: The sample points for the simulation networks generated using the Gaussian noise $N(0, 0.05^2)$  on the base network.}}
	\label{Fig:Config_ARC}
\end{figure}

We used a different simulation setting for comparing the proposed method against existing methods. We constructed topologically different shapes by combining circular arcs with and without a gap (Figure \ref{Fig:Config_ARC}). The networks were generated by sampling points from different topological shapes. We considered three topologically different networks with the difference in their number of loops in each group. Group 1 has three loops, Group 2 has two loops and Group 3 has one loop. An individual network in each group is generated by first sampling the coordinates $y_i$ for the $i$-th node along the ground truth patterns. The coordinates $y_i$ are perturbed with  Gaussian noise $N(0, 0.05^2)$. The  weight $w_{ij}$ between nodes $y_i$ and $y_j$ is given by the Euclidean distance. To retain only the dominant loops in the network, we applied the following thresholding scheme
$$
w_{ij}^\prime = w_{ij} (1-I_{ij} ) + 10^{-3}   I_{ij} \cdot U(0, 1),
$$
where $U(0, 1)$ is the uniform distribution on the interval $(0, 1)$ and the indicator function $I_{ij} = 1$ if $w_{ij} > 0.5$ and $0$ otherwise. The edge weights $w_{ij}^\prime$ are constructed such that the connections larger than the threshold 0.5 are replaced with random noise to retain only the dominant loops in the networks. We generated $60$ random networks per group.

\begin{table}[h!]
	\caption{The performance results showing average p-values with the standard deviations. The false positive and false negative rates are shown in the brackets. Smaller error rates are preferred. The graph theory features Q-modularity  $\mathfrak{L}_{Q}$ and betweenness $\mathfrak{L}_{bet}$ are used. The Gromov-Hausdorff distance $\mathfrak{L}_{GH}$ and  the bottleneck distance $\mathfrak{L}_{BN}$ in persistent homology are used. The $\mathfrak{L_w}$ is the proposed Wasserstein distance  on death values. $\mathfrak{L_c}$ is the proposed test statistic on the common 1-cycle basis.}
	\label{Table:SimulationStudy_arc}
	\centering
	\begin{tabular}{c|c|c|c|c|c|c}
		Groups & $\mathfrak{L}_{Q}$ & $\mathfrak{L}_{bet}$ & $\mathfrak{L}_{GH}$ &  $\mathfrak{L}_{BN}$ &  $\mathfrak{L_w}$ &  $\mathfrak{Lc}$ \\ \hline
		
		\multirow{2}{*}{ 1 vs. 2} & $0.4192$ & $0.5167$ & $0.4880$ & $0.4495$ & $0.0023$ & $0.0000$ \\
		
		& $\pm 0.28$ & $\pm 0.28$  & $\pm 0.30$ & $\pm 0.30$ & $\pm 0.00$ & $\pm 0.00$\\ 
		
		& $(0.94)$ & $(0.98)$  & $(0.88)$ & $(0.92)$ & $(0.00)$ & $(0.00)$\\ \hline
		
		\multirow{2}{*}{ 1 vs. 3} & $0.3521$ & $0.4673$  & $0.4903$ & $0.5419$ & $0.0000$ & $0.0000$\\
		
		& $\pm 0.31$ & $\pm 0.27$  & $\pm 0.29$ & $\pm 0.30$ & $\pm 0.00$ & $\pm 0.00$\\ 
		
		& $(0.76)$ & $(0.98)$ & $(0.94)$ & $(0.98)$ & $(0.00)$ & $(0.00)$\\ \hline
		
		\multirow{2}{*}{ 2 vs. 3} & $0.5671$ & $0.4672$ & $0.5503$ & $0.4362$ & $0.0144$ & $0.0000$\\
		
		& $\pm 0.28$ & $\pm 0.30$  & $\pm 0.29$ & $\pm 0.29$ & $\pm 0.03$ & $\pm 0.00$\\
		 
		& $(0.96)$ & $(0.96)$ & $(0.98)$ & $(1.00)$ & $(0.06)$ & $(0.00)$\\ \hline \hline
		
		\multirow{2}{*}{ 1 vs. 1} &  $0.5399$ & $0.5170$ & $0.4251$ & $0.4793$  & $0.5069$ & $0.4917$ \\
		
		& $\pm 0.26$ & $\pm 0.28$  & $\pm 0.26$ & $\pm 0.27$ & $\pm 0.29$ & $\pm 0.25$\\ 
		
		& $(0.04)$ & $(0.04)$ & $(0.08)$ & $(0.04)$ & $(0.04)$ & $(0.06)$ \\ \hline
		
		\multirow{2}{*}{ 2 vs. 2} & $0.5487$ & $0.5291$ & $0.5153$ & $0.5031$ & $0.4524$ & $0.5164$\\
		
		& $\pm 0.30$ & $\pm 0.26$  & $\pm 0.29$ & $\pm 0.30$ & $\pm 0.26$ & $\pm 0.32$\\ 
		
		& $(0.02)$ & $(0.02)$ & $(0.04)$ & $(0.04)$ & $(0.04)$ & $(0.14)$\\ \hline 
		
		\multirow{2}{*}{ 3 vs. 3} & $0.4836$ & $0.4608$ & $0.5322$ & $0.5464$ & $0.5069$ & $0.5086$ \\
		
		& $\pm 0.30$ & $\pm 0.26$  & $\pm 0.25$ & $\pm 0.33$ & $\pm 0.32$ & $\pm 0.31$\\
		 
		& $(0.08)$ & $(0.06)$ & $(0.04)$ & $(0.10)$ & $(0.04)$ & $(0.04)$\\ \hline
		
	\end{tabular}
\end{table}

We compared our model to graph theory features (Q-modularity, betweenness) \cite{rubinov2010complex} and persistent homology methods (Gromov-Hausdorff and bottleneck distances) \cite{lee.2011.MICCAI,chung.2019.NN}. Table \ref{Table:SimulationStudy_arc} shows the performance results with the average p-values with the standard deviations. The false negative rates and false positive rates are also given in the brackets. $\mathfrak{L}_{Q}$ and $\mathfrak{L}_{bet}$ are based on the Q-modularity and betweenness. For the graph theory features, we used a similar test statistic  as (\ref{Eq:coefflossdef}) and used the maximum absolute difference in the average graph theory features in each group.  $\mathfrak{L}_{GH}$ and  $\mathfrak{L}_{BN}$ are based on the Gromov-Hausdorff  and bottleneck distances. The computation of the Gromov-Hausdorff and bottleneck distances follows the methods in \cite{chung.2019.NN}. $\mathfrak{L_w}$ is the proposed Wasserstein distance based on death values. $\mathfrak{L_c}$ is the proposed statistical inference based on the common 1-cycle basis.	In all the test, we used the permutation test with 100000 permutations. 

In testing topological differences (first three rows), the existing methods did not perform well failing to identify the topological differences. The proposed methods $\mathfrak{L_w}$ and $\mathfrak{L_c}$ performed very well and were able to differentiate topological differences. In testing {\em no} topological difference (last three rows), all the methods performed reasonably well and did not report any false positives. If there are subtle topological differences that are difficult to differentiate,  existing methods will likely to  fail while the topological method will likely to detect signals. 

\section{Application}
\subsection{Dataset and preprocessing}
\label{sec:dataset}
In this study, we used the the subset of the resting-state fMRI data collected in the Human Connectome Project (HCP) \cite{van2012human,glasser2013minimal}. We used the data used in \cite{huang2020statistical}. The subjects were in ages ranging from 22 to 36 years with average age $29.24 \pm 3.39$ years for 172 males and 240 females. The fMRI data were acquired for approximately 15 minutes for each scan. The participants are at rest with eyes open with relaxed fixation on a projected bright cross-hair on a dark back-ground \cite{van2012human}. The fMRI data were collected on a customized Siemens 3T Connectome Skyra scanner using a gradient-echoplanar imaging (EPI) sequence with multiband factor 8, repetition time (TR) $720 ms$, time echo (TE) $33.1 ms$, flip angle $52^\circ$, $104 \times 90$ (RO $\times$ PE) matrix size, 72 slices, $2 mm$ isotropic voxels, and 1200 time points is used. 

The standard minimal preprocessing pipelines \cite{glasser2013minimal} such as spatial distortion removal \cite{andersson2003correct}, motion correction \cite{jenkinson2001global}, bias field reduction \cite{glasser2011mapping}, registration to the structural MNI template, and data masking using the brain mask obtained from FreeSurfer \cite{glasser2013minimal} is performed on the fMRI scans. This resulted in the resting-state fMRI  with $91 \times 109 \times 91$,  $2mm$ isotropic voxels at 1200 time points.

The scrubbing is done to remove fMRI volumes with spatial artifacts in functional connectivity \cite{power2012spurious} due to 
significant head motion \cite{power2012spurious,huang2020statistical}. The framewise displacement (FD) from the three translational displacements and three rotational displacements at each time point to measure the head movement from one volume to the next is calculated. The volumes with FD larger than $0.5mm$ and their neighbors were scrubbed \cite{power2012spurious,huang2020statistical}. $12$ subjects having excessive head movement are excluded from the dataset, resulting in a refined fMRI dataset of 400 subjects (168 males and 232 females), which we analyzed in this study. 
Subsequently, the Automated Anatomical Labeling (AAL) template was applied to parcellate the brain volume into 116 non-overlapping anatomical regions \cite{tzourio2002automated}. The fMRI across voxels within each brain parcellation is averaged, which  resulted in 116 averaged fMRI time series with 1200 time points for each subject. The additional details on the processing can be found in \cite{huang2020statistical,song.2021.MICCAI}. 

\begin{figure*}[t!]
	\centering
	\includegraphics[width=0.7\textwidth]{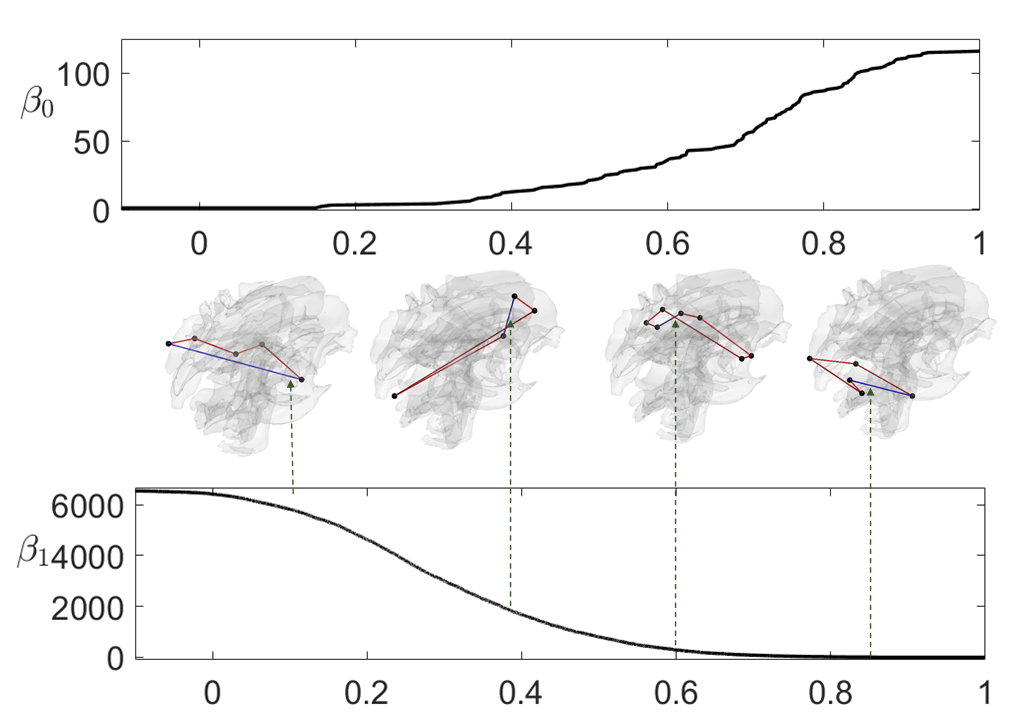}
	\caption{The graph filtration of the average correlation network of 400 subjects. $\beta_0$ is monotonically increasing while $\beta_1$ is monotonically decreasing over the filtration. We have total $6555$ cycles in the brain network. Middle: Four 1-cycles chosen at specific death values are shown. The edges that destroy the cycles are shown in blue.}
	\label{fig:braindata}
\end{figure*}

\subsection{Cycle computation}

For each subject, we measured the whole-brain functional connectivity by computing the Pearson correlation matrix $\rho = (\rho_{ij})$ over while time points across 116 brain regions. This resulted  in 400 correlation matrices of size $116 \times116$. Since the dataset contains $p=116$ nodes, the total number of edges in the brain network is $q = p(p-1)/2 = 6670$.

We then performed the birth-death decomposition following Theorem \ref{Th:bddecomp}. The number of edges in the birth set is 
$$\mathcal{P} = p - 1 = 116-1 = 115.$$ The number of edges in the death set is $$\mathcal{Q} = q - \mathcal{P} = 6555.$$ The edges from the death set are then sequentially added to the birth set to generate a sequence of 6555 subnetworks. Each subnetwork has only one cycle which is identified using the Hodge Laplacian. Figure \ref{fig:braindata} displays how the number of the topological invariants $\beta_0$ (number of connected components) and $\beta_1$ (number of cycles) change over the graph filtration. $\beta_0$ remains at one for a long duration and  begins to increase after correlation value 0.4 and eventually reaches 116 which is the number of independent components or nodes. On the other hand, $\beta_1$ begins with $\mathcal{Q} = 6555$ number of cycles and keeps decreasing as the edges are removed sequentially and eventually reaches near zero after correlation 0.6. The correlation between 0.4 and 0.6 is the range where the topological structure of the brain networks seem to change. Once all the cycles are identified and extracted, we mainly focused on the death values of cycles. These topological quantities are used as test statistics for discriminating males from females.

  \begin{figure*}[t!]
	\centering
	\includegraphics[width=1\textwidth]{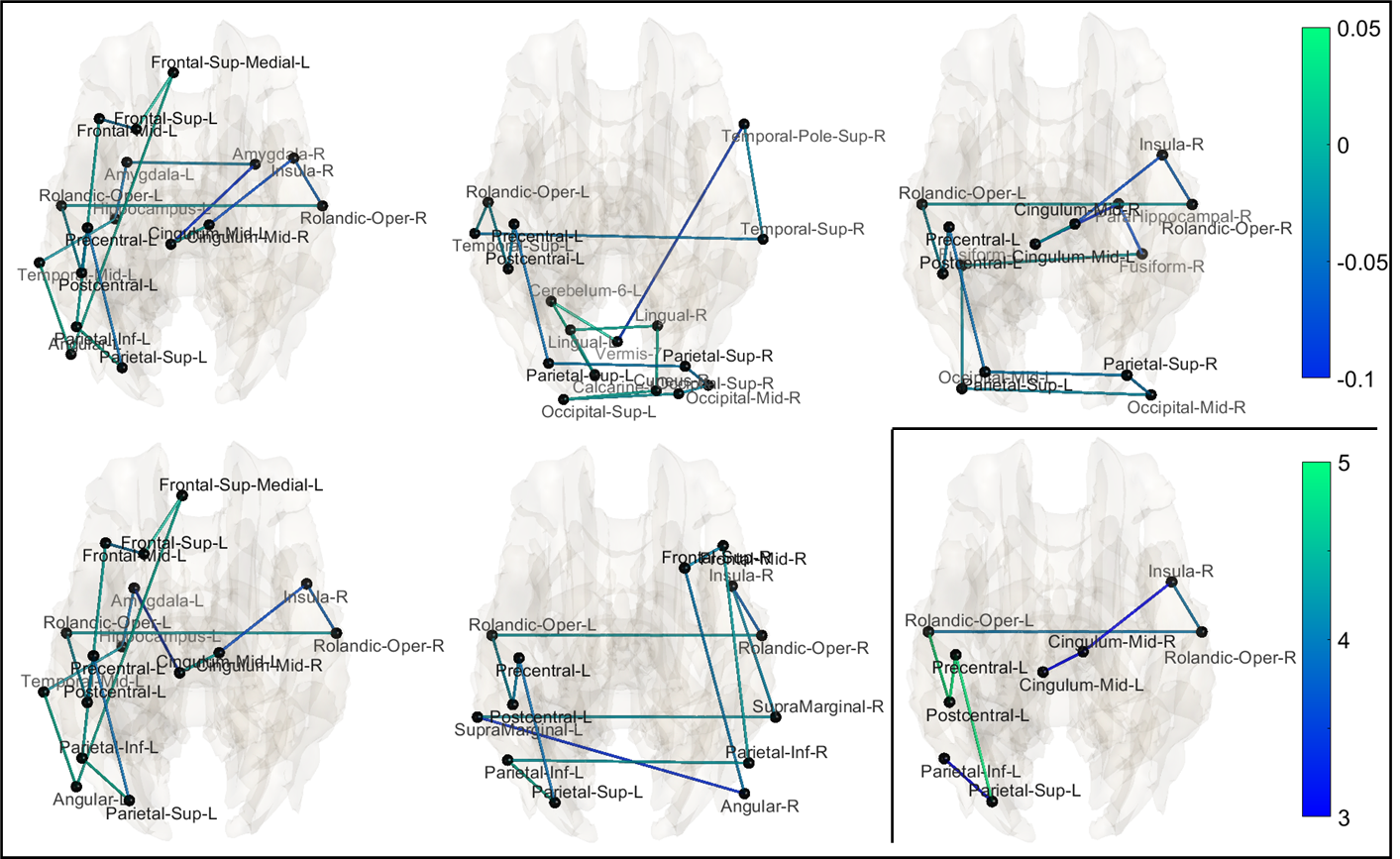}
	\caption{The five most discriminating cycles having maximum values in the test statistics  are shown. The color bar  shows the difference between the average correlations  (female - male)   for the five cycles. Bottom Right: The edges that frequently occur in all the five cycles are shown. The color bar is the overlap frequency.}
	\label{Fig:FiveMostDiscCycles}
\end{figure*}

\subsection{Inference on cycle differences}
\subsubsection{Death values} The topological similarity between the networks can be measured by computing the 2-Wasserstein distance between persistent diagrams of 1-cycles \cite{song.2021.MICCAI}. We computed the average within and between group distances. Then the transposition test   \cite{chung2019rapid} on the ratio statistic was carried out in determining the statistical significance between 232 female and 168 male brain networks. The  observed test statistic is 1.0232 and corresponding p-value is 0.049 based on 500000 random transpositions.

\subsubsection{Common 1-cycle basis}
The common 1-cycle basis was obtained from the  average correlation matrices of  400 subjects.
We computed the coefficients for each network using  (\ref{Eq:kcyclecoeff}) and the mean coefficients for females and males separately for each cycle. We then used the maximum difference (\ref{Eq:coefflossdef}) between mean coefficients as the test statistic. The observed statistic was  0.408, which corresponds to the p-value of 0.03 based on  500000 permutations.

The five most discriminating cycles are identified by identifying which cycle gives the  the maximum test statistic values in the decreasing order: $0.408$, $0.407$, $0.405$, $0.396$ and $0.393$ corresponding to the cycle indexing 2446, 1140, 4090, 3683 and 831. Figure \ref{Fig:FiveMostDiscCycles} shows five most discriminating cycles corresponding to the maximum observed statistics. Some connections consistently appear in all the five cycles.  The nodes observed in the five most discriminating cycles are superior parietal gyrus (Parietal-Sup-L), inferior parietal lobule (Parietal-Inf-L), Precentral gyrus (Precentral-L),  Postcentral gyrus (Postcentral-L), the rolandic operculum (Rolandic-Oper-L, Rolandic-Oper-R), the median cingulate and para cingulate gyri (Cingulum-Mid-R, Cingulum-Mid-L) and the Insula. The connections between these regions highlight their importance in discriminating males and females. The symmetric connection between the left and right rolandic operculum, superior parietal lobule and the middle cingulate appear in at least $3$ most dominating cycles. We determined the overlap frequency  and displayed in the bottom right of Figure \ref{Fig:FiveMostDiscCycles}, where the color scale correspond to the the number of overlaps in the $5$ cycles. 

There is known sex difference in the parietal region, which is involved in spatial ability such as the mental rotation \cite{koscik.2009}. \cite{xu.2015} reported sex differences in the left parietal, precentral and  postcentral regions in a rs-fMRI study, where Kendall's coefficient of concordance (KCC) was used to measure the similarity of the ranked time series of a given voxel to its nearest 26 neighbor voxels \cite{zang.2004}. The sex difference is reported in the left rolandic operculum in rs-fMRI study \cite{rubin.2017}. While all these pervious studies are reporting the sex differences at the node level, we are consistently identifying them within 5 most dominant cycles. The edges connecting Rolandic-Oper-L, Rolandic-Oper-R and Insula appear in $4$ cycles. The edges connecting Parietal-Sup-L and Parietal-Inf-L and the edges connecting Cingulum-Mid-R, Cingulum-Mid-L and Insula-R occur in $3$ cycles. We believe these brain regions can act as discriminating biomarkers for sexual dimorphism studies including Alzheimer's disease which affects disproportionately more women than men \cite{fisher.2018}.

\subsection{Comparison against baselines}
We compared the discriminating power of our method against  Gromov-Hausdorff (GH) and bottleneck (BN) distances often used in persistent homology. The computed p-values are 0.540 for GH and 0.277 for BN and not able to discriminate the networks. Both the GH and BN distances did not perform well in the real data. We also used graph theory features Q-modularity and betweenness   and obtained  p-values of 0.035 and 0.6202 respectively \cite{rubinov2010complex}. Among all 4 baseline methods, Even though Q-modularity performed well, it cannot be used to identify connections that are responsible for the differences and explicitly localize regions that cause significant topological disparity.

\section{Conclusion}
Cycles in the brain network are one of the most fundamental topological features  in understanding higher order interactions. In this study, an efficient scalable algorithm to identify and extract the 1-cycles in a network is proposed. We combine the ideas from persistent homology and the Hodge Laplacian in developing the spectral version of topological data analysis. The proposed spectral-TDA  is demonstrated with an illustration and applied to the resting state brain networks from Human Connectome Project (HCP). The proposed algorithm is efficient for typical brain network data which has few hundred nodes ($p \sim 100$). Even for  larger networks ($p \gg 1000$), various computations can be done quickly in $\mathcal{O}(p \log p)$ run time using the birth-death decomposition \cite{songdechakraiwut2020topological}. 

One of our major goals in the study is to discriminate networks having different cycles. It is not even clear how to algebraically represent cycles. To capture this topological characteristic of loops, we used the common 1-cycle basis to precisely encode this information across subjects. Through the combination of MST and the Hodge Laplacian, we were able to extract and represent 1-cycle basis as a sparse matrix. Any cycle in the graph is represented as a linear combination of basis
$ \sum_{j=1}^Q \alpha_j \mathcal{C}^j.$
Such a vectorization enables us to build more complex models such as sparse network models  or joint identification of common cycles across subjects \cite{lee2014hole}. This is left as a future study.

We designed a new topological inference procedure based on the 1-cycle attributes such as death values of cycles.  
 The Wasserstein distance between cycles $\mathcal{C}^i$ and $\mathcal{C}^j$ is simply the squared difference of death values $(d_i -d_j)^2$. Such squared norm makes computations involving cycles straightforward. 
 The new Wasserstein distance  based statistical framework is used in discriminating the brain networks of males and females. Our study emphasizes that it is meaningful to study and model the higher order interactions using cycles for brain network analysis.

\section*{Acknowledgment}
We thank Kelin Xia of Nanyang Technological University for discussion and support of the project. 
We also like to thank Shih-Gu Huang of National University of Singapore for providing support for fMRI processing. 
We also thank Sixtus Dakurah and Soumya Das of University of Wisconsin-Madison for discussion on the statististical analysis.

\bibliographystyle{ieeetr}
\bibliography{reference.2022.08.13moo,reference.2021.08.18,reference.2021.10.08} 

\end{document}